\documentclass[prb,twocolumn,showpacs,superscriptaddress]{revtex4}

\usepackage{graphicx}
\usepackage{dcolumn}
\usepackage{bm}

\begin{document}

\title{Phase-charge duality in Josephson junction circuits: Role of inertia and effect of microwave irradiation}
\author{W. Guichard}
\affiliation{Institut N\'eel, C.N.R.S. and Universit\'e Joseph Fourier, BP 166, 38042 Grenoble, France}
\author{F. W. J. Hekking}
\affiliation{Universit\'e Joseph Fourier, Laboratoire de Physique et Mod\'elisation des Milieux Condens\'es, C.N.R.S., BP 166, 38042 Grenoble, France}
\date{\today }

\begin{abstract}
We investigate the physics of coherent quantum phase slips in two distinct circuits
containing small Josephson junctions: (i) a single junction embedded in an inductive
environment and (ii) a long chain of junctions. Starting from the standard Josephson
Hamiltonian, the single junction circuit can be analyzed using quasi-classical methods;
we formulate the conditions under which the resulting quasi-charge dynamics is exactly
dual to the usual phase dynamics associated with Josephson tunneling. For the chain we
use the fact that its collective behavior can be characterized by one variable: the
number $m$ of quantum phase slips present on it. We conclude that the dynamics of the
conjugate quasi-charge is again exactly dual to the standard phase dynamics of a single
Josephson junction. In both cases we elucidate the role of the inductance, essential to
obtain exact duality. These conclusions have profound consequences for the behavior of
single junctions and chains under microwave irradiation.  Since both systems are governed
by a model exactly dual to the standard resistively and capacitively shunted junction
model, we expect the appearance of current-Shapiro steps. We numerically calculate the
corresponding current-voltage characteristics in a wide range of parameters. Our results
are of interest in view of a metrological current standard.
\end{abstract}

\maketitle

\section{Introduction}

Two physical systems that can be mapped onto each other by interchanging the role of
position and its canonically conjugate momentum are said to be related by duality. If the
physical properties of one of the systems are known, those of its dual counterpart can be
predicted by applying the set of duality transformations that accompany the position and
momentum interchange and relate the parameters of the two systems. In some special cases,
duality maps the system onto itself; one then speaks of self-duality. An example of a
system that exhibits exact self-duality is the harmonic oscillator. More frequently one
encounters systems that exhibit an approximate self-duality relating the system's
asymptotic behavior in two different limiting parameter regimes.

\begin{figure}
\centering
\includegraphics[width=\columnwidth,angle=0]{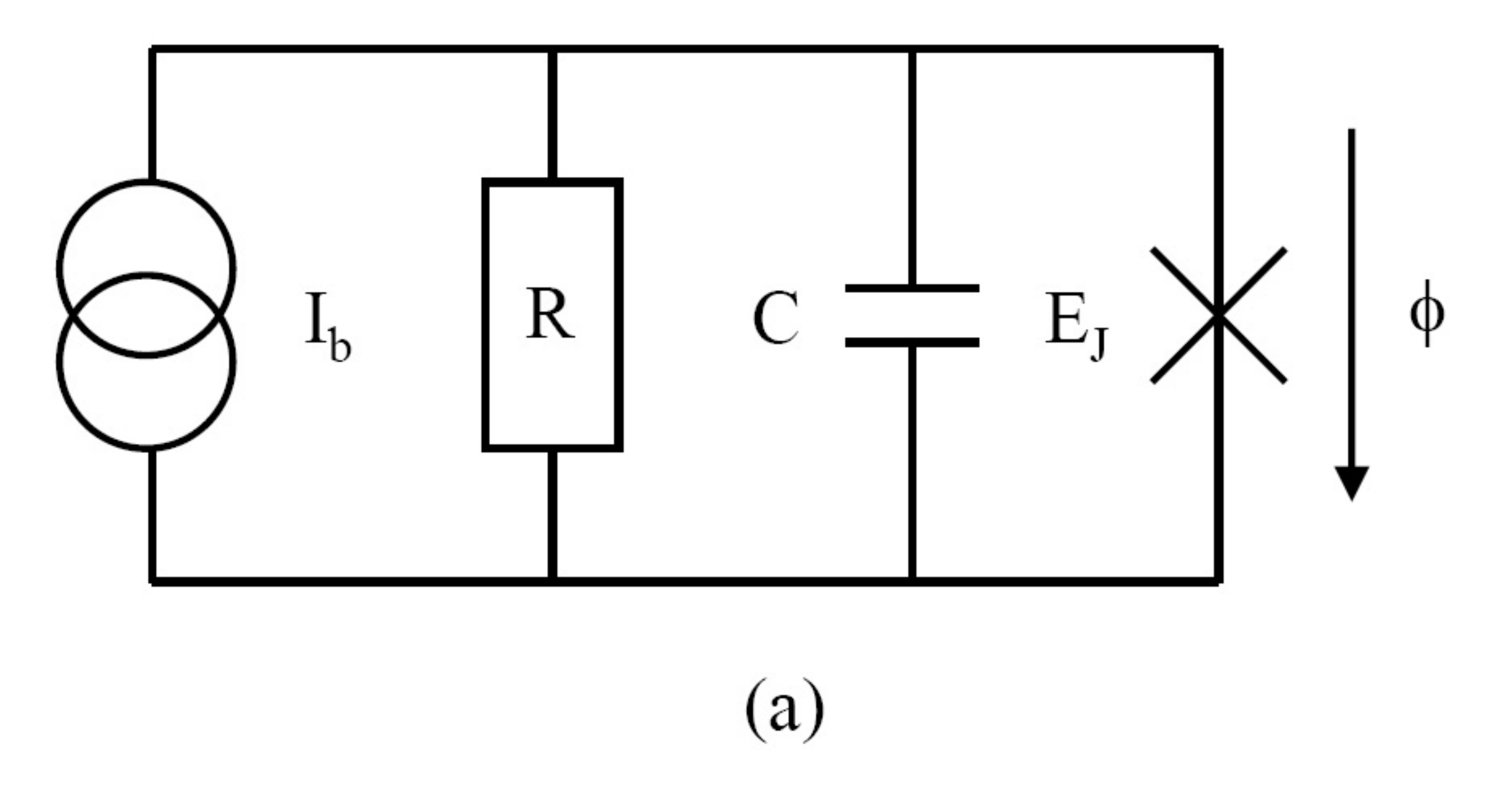}
\includegraphics[width=\columnwidth,angle=0]{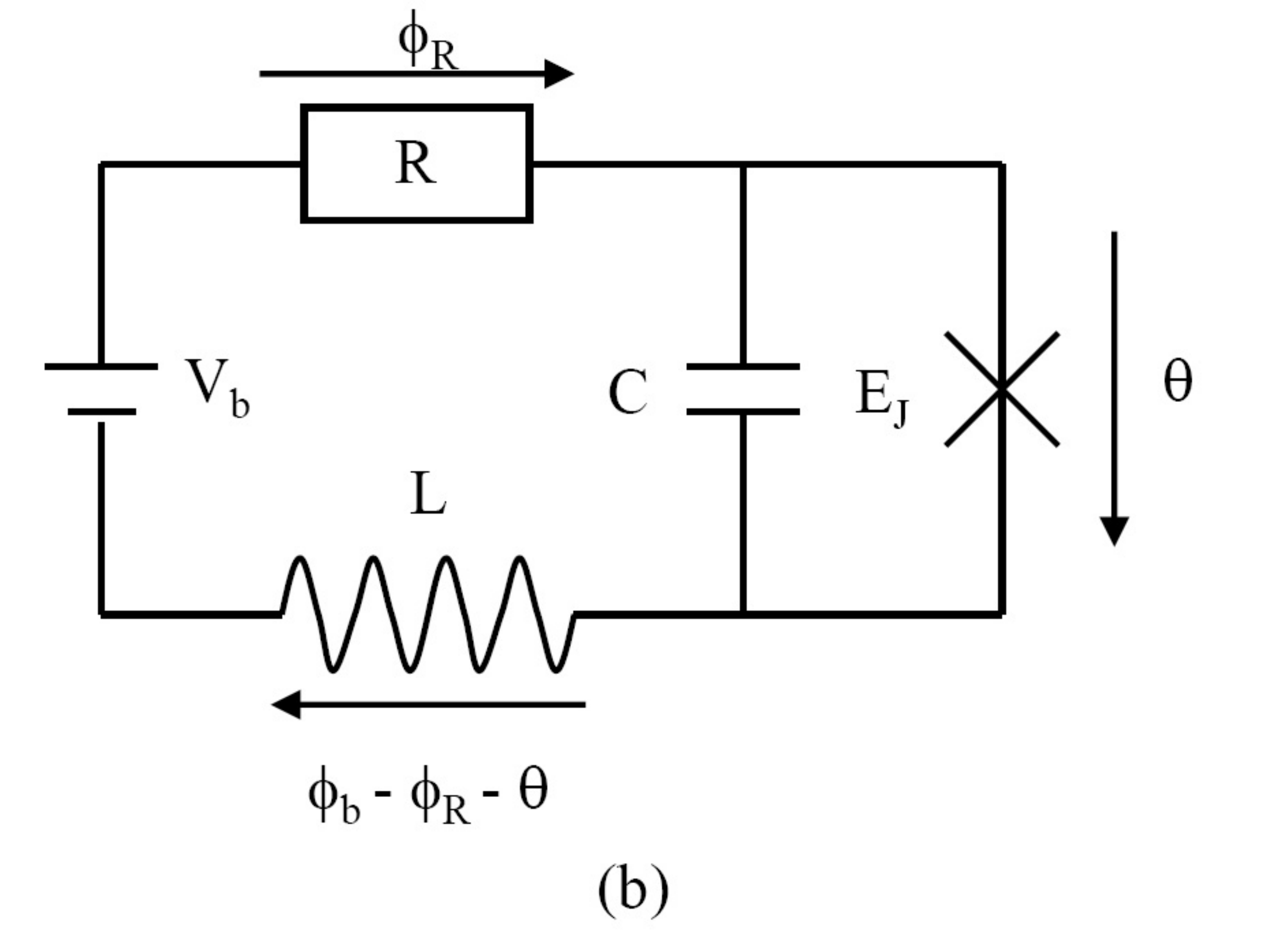}
\caption{(a) RCSJ-model: Current-biased Josephson junction (cross), bias current $I_b$,
with Josephson energy $E_J$ shunted by a capacitor $C$ and a resistor $R$. (b)
CJRL-model: Voltage-biased Josephson junction (cross), bias voltage $V_b$, shunted by a capacitor $C$ and in series with
an inductor $L$ and a resistor $R$.}\label{JJ}
\end{figure}

Duality transformations have been proven useful in a variety of situations from a broad
range of fields~\cite{Savit_80} including statistical mechanics, condensed matter physics
and gauge-field theories. Here we will focus on the case of Josephson junctions embedded
in an electromagnetic environment, where the duality associated with the conjugate charge
and phase degrees of freedom has been employed to study the circuit's dissipative
dynamics~\cite{Schmid_83,SchoenZaikin_90,Weiss_99,Ingold_99}. A Josephson junction, as
depicted schematically for a current-biased situation in Fig.~\ref{JJ}a, is characterized
by two competing energy scales: the Josephson coupling energy $E_J = \hbar I_c/2e$,
proportional to the Josephson critical current $I_c$, and the charging energy $E_C=
e^2/2C$, inversely proportional to the junction's capacitance $C$. A self-duality
property can be used to relate the junction's behavior in the presence of a resistor $R$
in the two limiting cases $E_J/E_C \gg 1$ and $E_J/E_C \ll
1$~\cite{Schmid_83,SchoenZaikin_90,Weiss_99,Ingold_99}. This self-duality is only
approximate, though; as we will discuss in more detail below, an exact duality
transformation exists between the circuit of Fig.~\ref{JJ}a and a different
superconducting circuit containing a large junction together with an additional element:
an inductance $L$~\cite{SchoenZaikin_90, Apenko_89}, see Fig.~\ref{JJ}b. This is to be
expected somehow: it is customary to describe the dynamics of the current-biased circuit
depicted in Fig.~1a in terms of a fictitious phase-particle of mass $C$. The dual
situation Fig.~\ref{JJ}b would then correspond to a voltage-biased circuit, the dynamics
of which is that of a charge-particle of mass $L$.

\begin{figure}
\centering
\includegraphics[width=0.9\columnwidth,angle=0]{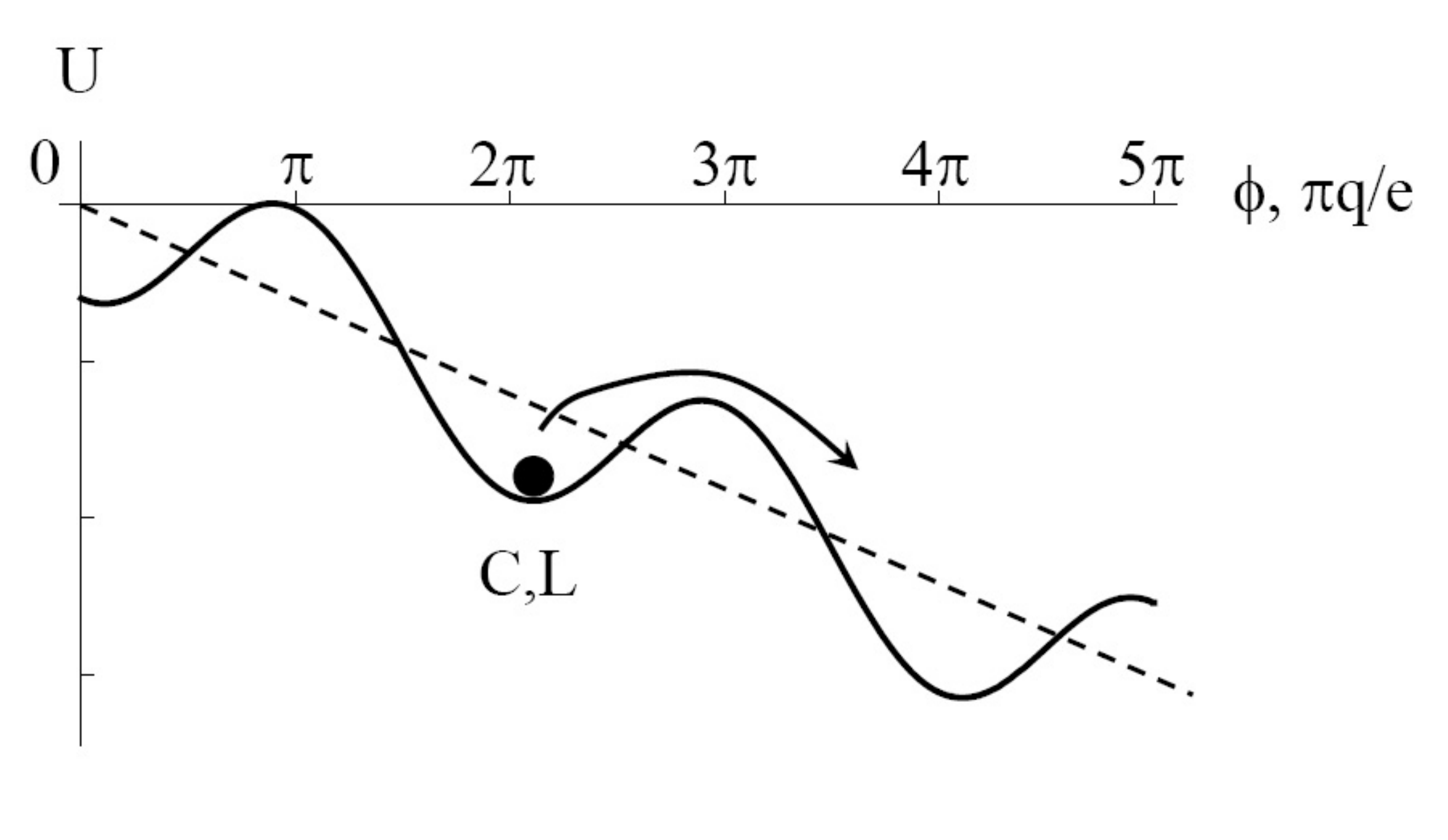}
\caption{Example of the tilted washboard potential $U$ as a function of phase $\phi$ and
charge $\pi q/e$ for a phase-particle of mass $C$ and a charge-particle of mass $L$,
respectively. The dashed line denotes the average tilt, proportional to the bias current
(phase-particle) through or the bias voltage (charge-particle) over the junction.}
\label{washboard}
\end{figure}

Let us push the duality analysis a little further. The phase-particle for the
current-biased circuit Fig.~\ref{JJ}a moves in a potential $U(\phi)$ which is the sum of
a periodic part and a linear tilt, see Fig.~\ref{washboard} for an example. The periodic
potential induces the tunneling of Cooper pairs of charge $2e$, its amplitude $E_J =
\hbar I_c/2e$ sets the maximum supercurrent $I_c$ that can be carried by the junction.
The tilt of the potential is proportional to the bias current $I_b$ through the junction.
By duality, the charge-particle for the voltage-biased circuit Fig.~\ref{JJ}b moves in a
potential $U(q)$ which also has a periodic part and a linear tilt, see
Fig.~\ref{washboard}. Here the periodic part induces a tunneling of phase or a "phase
slip" whereby the phase difference across the junction winds by an amount $2\pi$;
accordingly, one can speak of a phase-slip junction. A winding of the phase with time
gives rise to a voltage over the element, and the amplitude of the periodic part of the
potential sets the maximum voltage $V_c$ the phase-slip element can sustain. The linear
tilt is proportional to the bias voltage $V_b$ across the junction.

Duality thus implies that the $I-V$ characteristics of the voltage-biased circuit in
Fig.~\ref{JJ}b can be obtained from those of the current-biased circuit of
Fig.~\ref{JJ}a, by simply interchanging the role of current and voltage. This has been
verified experimentally in Ref.~\cite{Corlevi_06}, where the measured $I-V$
characteristics of an underdamped Josephson junction were found to be exactly dual to
those of the overdamped Josephson junction measured in Ref.~\cite{Steinbach_01}, in
accordance with the approximate self-duality exhibited by a Josephson junction in these
respective limits.

Probably one of the most important consequences of duality in this connection is the
case of a junction that is irradiated by microwaves (MW) of frequency $f$. If the MW
frequency $f$ is commensurate with the frequency of the motion of the phase-particle
in the periodic potential, phase-locking occurs yielding Shapiro steps~\cite{Shapiro_63},
in the $I-V$ characteristics at well-defined voltages that are proportional to
multiples $n$ of the applied frequency $f$: $V_n = n h f/2e$, where $h$ is Planck's
constant and $e$ the elementary charge. As frequency can be controlled with an extreme
accuracy, this effect is currently used in metrology to define the voltage standard,
for a review see~\cite{Kautz_96, metrology}. Observation of the dual phenomenon
--- phase locking for the charge-particle yielding Shapiro steps at well-defined currents
that are multiples of the applied frequency $I_n = n 2e f$ --- would have far-reaching
consequences for metrology, as this would enable one to define a current standard with an
unprecedented precision.

Clearly, the voltage-biased circuit illustrated in Fig.~\ref{JJ}b is not the only one
dual to that of Fig.~\ref{JJ}a. In fact, {\em any} circuit element that features
appropriate phase tunneling is a possible candidate for duality. In view of experimental
implementations and applications it is interesting to compare several possibilities. In a
recent paper, Mooij and Nazarov~\cite{Mooij_06} proposed exact duality between
Fig.~\ref{JJ}a and a voltage-biased circuit containing a narrow superconducting
wire~\cite{Zaikin_97,Arutyunov_08}. As we will detail below another possibility would be to use a
one-dimensional chain of Josephson junctions.

The paper is organized as follows. For pedagogical reasons and also for the sake of
completeness, we start by reviewing the physics of the circuits of Fig.~\ref{JJ} in
Section~\ref{charge_phase}. This enables us to demonstrate the duality principle as well
as its consequences for superconducting circuits with the aid of a relatively simple
example. In Section~\ref{arrays} we demonstrate that a voltage-biased chain of junctions
is exactly dual to Fig.~\ref{JJ}a. The current-voltage characteristics of phase-slip
junctions both in the absence and in the presence of MW irradiation are discussed in
Section~\ref{MW_effects}; some perspectives and experimental consequences of our
theoretical study are discussed in Section~\ref{outlook}.

\section{Charge-phase duality in circuits containing a single junction}
\label{charge_phase}

In this Section we wish to demonstrate that the two superconducting circuits depicted
in Fig.~\ref{JJ} are dual to each other. Specifically, we will establish the
conditions under which this duality holds, treating both circuits entirely
quantum mechanically. We will discuss the requirements to be met such that the classical limit can be
taken, leading to the usual resistively and capacitively shunted junction (RCSJ) model~\cite{Tinkham} for Fig.~\ref{JJ}a and its dual counterpart,
the capacitively shunted junction in series with a resistor and an inductance (CJRL) model for Fig.~\ref{JJ}b.

\subsection{Capacitively shunted junction: phase-inertia}
We start our analysis by considering the circuit shown in Fig.~\ref{JJ}a. It contains a
current-biased Josephson junction (bias current $I_b$), shunted by a capacitance $C$ and
a resistance $R$. Let $\hat{\phi}$ be the operator corresponding to the phase difference
across the junction and $\hat{Q}$ the canonically conjugate charge, such that the
commutator $[\hat{Q},\hat{\phi}] = -2ie$. The resistor induces dissipation that we will
account for within the framework of the Caldeira-Leggett
model~\cite{Caldeira_83,SchoenZaikin_90,IngoldNazarov,Weiss_99}. Hence the circuit
presented in Fig.~\ref{JJ}a can be described by the Hamiltonian $\hat{H} = \hat{H}_0 +
\hat{H}_B$ where
\begin{eqnarray}
\hat{H}_0 = \frac{\hat{Q}^2}{2C} + U(\hat{\phi}) \mbox{ , } U(\hat{\phi}) = - E_J \cos \hat{\phi} - \hbar I_b \hat{\phi}/2e, \\
\hat{H}_B = \sum_{i=1}^\infty \frac{\hat{P}_i^2}{2} + \frac{\omega_i^2}{2} (\hat{X}_i -
\frac{c_i}{\omega_i^2}\hat{\phi})^2 .
\end{eqnarray}
Here $\hat{H}_B$ is the Caldeira-Leggett Hamiltonian describing a bath of oscillators
with frequencies $\omega_i$, conjugate momenta and positions $\hat{P}_i$ and
$\hat{X}_i$, the latter coupling linearly to the junction's phase operator
$\hat{\phi}$ with coupling constants $c_i$.

From the Hamiltonian $\hat{H}$, one can
obtain the equation of motion for the operator $\hat{\phi}$,
\begin{equation}
\hbar C \ddot{\hat{\phi}}/2e +  I_c \sin\phi = I_b + \delta \hat{I}, \label{preRCSJ}
\end{equation}
where the current $\delta \hat{I}$ is related to the momenta of the oscillator bath,
\begin{equation}
\delta \hat{I} = -\frac{2e}{\hbar} \sum_{i=1}^\infty  \frac{c_i}{\omega_i^2}
\dot{\hat{P}}_i. \label{dI}
\end{equation}
The bath momenta satisfy the equation of motion
\begin{equation}
\ddot{\hat{P}}_i + \omega_i^2 \hat{P}_i = c_i \dot{\hat{\phi}}. \label{P_i}
\end{equation}
Direct integration of Eq.~(\ref{P_i}) and substitution of the result into Eq.~(\ref{dI})
yields
\begin{equation}
\delta \hat{I} = \hat{i}(t) - \int \limits_0 ^t Y(t-t') \hbar \dot{\hat{\phi}}(t')/2e,
\end{equation}
where the first term $\hat{i}(t)$ is related to the homogeneous solution of
Eq.~(\ref{P_i}); it is random in nature due to the uncertainty with respect to the bath's
initial conditions. The second term is related to the particular solution of
Eq.~(\ref{P_i}); it describes the response of the bath to the voltage operator $\hbar
\dot{\hat{\phi}}/2e$ through the retarded admittance $Y(t)$ with Fourier transform
\begin{equation}
Y(\omega) = \left(\frac{2e}{\hbar}\right)^2\sum _{i=1}^\infty
\frac{c_i^2}{\omega_i^2}\frac{i \omega}{(\omega +i \eta)^2 - \omega_i^2}.
\end{equation}
Provided we choose the bath parameters $c_i$ and $\omega_i$ such that
\begin{equation}
\Re \mbox{e}[Y(\omega)] = \left(\frac{2e}{\hbar}\right)^2 \pi \sum \limits_{i=1}^\infty
\frac{c_i^2}{\omega_i}\delta(\omega^2 - \omega_i^2) = 1/R,
\end{equation}
the bath's response is ohmic corresponding to that of a resistance $R$. As a result we
can present Eq.~(\ref{preRCSJ}) in the form
\begin{equation}
\hbar C \ddot{\hat{\phi}}/2e +  \hbar \dot{\hat{\phi}}/2eR + I_c \sin\hat{\phi} = I_b +
\hat{i} \label{RCSJ}.
\end{equation}
For later use, it is convenient to write Eq.~(\ref{RCSJ}) in a dimensionless form, dividing both sides by
$I_c$; one then obtains
\begin{equation}
d^2 \hat{\phi}/d\tau^2 +  \sigma d\hat{\phi}/d\tau +  \sin\hat{\phi} = \bar{I}_b +
\hat{\bar{i}} \label{RCSJdimless},
\end{equation}
where $\tau = \omega_p t$ with $\omega_p = (8 E_J E_C)^{1/2}/\hbar$ the junction's plasma
frequency, $\sigma = (\hbar/2e I_c C)^{1/2}/R$ the dimensionless inverse resistance,
$\bar{I}_b = I_b/I_c$ and $\hat{\bar{i}} = \hat{i}/I_c$.

Equation~(\ref{RCSJ}) is a non-linear quantum Langevin equation, owing its stochastic nature to the presence of the random operator $\hat{i}$. The statistics of $\hat{i}$ will be fixed by assuming
the initial state of the bath to be the canonical equilibrium
one at temperature $T$. This, together with the harmonic nature of the bath, implies that
the statistics of the random operator $\hat{i}$ is Gaussian with
average value $\langle \hat{i} \rangle =0$. In accordance with the fluctuation-dissipation theorem, the symmetrized second moment
$\langle \{\hat{i} (t), \hat{i} (0)\}\rangle/2$, where $\{ \ldots, \ldots\}$ denotes the anti-commutator, is then characterized by the spectral function
\begin{eqnarray}
S_i(\omega) &\equiv& \int dt e^{i \omega t} \langle \{\hat{i} (t), \hat{i} (0)\}\rangle/2 \nonumber \\
 &=& \hbar \omega \Re \mbox{e}[Y(\omega)] \coth (\hbar \omega/2k_B T) \nonumber \\
 &=& \frac{\hbar \omega}{R} \coth (\hbar \omega/2 k_B T).
\label{symspecfunci}
\end{eqnarray}

A general analysis of the quantum Langevin equation (\ref{RCSJ}) is beyond the scope of
this article. Here we are interested in the classical limit of the operator equation
(\ref{RCSJ}) where it reduces to the well-known resistively and capacitively shunted
junction (RCSJ) model, describing the classical dynamics of a fictitious phase-particle.
The capacitor provides the particle's inertia; the corresponding acceleration is the
capacitor's displacement current. The resistor provides both velocity-proportional
damping $\sim \dot{\phi}/R$ and classical noise $i(t)$. The noise $i$ adds to the applied
bias current $I_b$ which, together with the junction's supercurrent $I_c \sin \phi$
yields the external force acting on the particle.

A classical interpretation of the operator equation (\ref{RCSJ}) makes sense if we can
accurately replace the operators by their respective expectation values, $\hat{\phi} \to
\phi$, $\hat{i} \to i$. In particular, we must be allowed to replace $\langle \sin
\hat{\phi} \rangle$ by $\sin \langle \hat{\phi} \rangle = \sin \phi$. For this to be
correct, the uncertainty $\delta \phi$ in the phase must be small compared to the period
of the sine function. The classical version of (\ref{RCSJ}) then describes the motion of
a narrow wave packet of width $\delta \phi \ll 1$. Such a wave packet can be constructed
as a superposition of extended phase states, implying an uncertainty on the level of the
junction charge $\delta Q$ that exceeds the elementary charge $e$. This means in
particular that the quasi-classical phase description does not capture effects associated
with Coulomb blockade~\cite{IngoldNazarov}.

It is useful to distinguish two cases, according the value of the parameter $\sigma$ in
Eq.~(\ref{RCSJdimless}): overdamped phase dynamics, corresponding to $\sigma
> 1$ and underdamped phase dynamics, $\sigma <1$.

In the overdamped case, the phase dynamics is always classical as damping times are
naturally short, of the order of $RC$. This is generally achieved in a low-resistance
environment that avoids charge localization. Classical phase dynamics also requires that
the environmental noise operator $\hat{i}$ can be treated
classically~\cite{Ivanchenko_69}. This corresponds to relatively high temperatures, such
that $S_i(\omega) = 2 k_B T/R$, see Eq.~(\ref{symspecfunci}). Then we can replace
$\hat{i}$ by a c-number $i$ such that $\langle i \rangle = 0 $ and $\langle i(t)i(0)
\rangle = 2 (k_B T/R) \delta (t)$; the noise is $\delta$-correlated. For this to be correct the
temperature $T$ should be large compared to the characteristic frequency $1/RC$. The
overdamped limit can be analyzed, e.g., by studying the Fokker-Planck
equation~\cite{Kampen} corresponding to the classical Langevin equation, as it was done
for the overdamped case in Ref.~\cite{Ivanchenko_69}.

In the underdamped case, damping times are long and wave packet spreading becomes
important. This issue is particularly relevant when the phase-particle is in the running
state corresponding to a finite voltage over the junction. In the absence of damping, the
uncertainty $\delta \phi$ can be kept within limits by applying classical time-dependent
external forces. The spreading of wave packets is governed by a rate proportional to the
kinetic energy. For the fictitious phase particle this corresponds to the charging energy
$E_C$, which is in fact the energy scale associated with charge localization leading to
Coulomb blockade. We require the time-dependence of the external force to be fast on the
scale $\hbar/E_C$. Physically this means that we work under conditions where charging
effects can be ignored. For example, in the case of externally applied microwaves (see
Section~\ref{MW_effects}) this implies their frequency to be larger than $E_C/\hbar$.

\subsection{Junction in series with an inductance: charge-inertia}
\label{juncind}

Next consider the circuit depicted in Fig.~\ref{JJ}b, where a single Josephson junction
is embedded in a combined inductive and resistive series environment (inductance $L$ and
resistance $R$)~\cite{Apenko_89,SchoenZaikin_90}. Its Hamiltonian is given by $\hat{H}' =
\hat{H}'_0 + \hat{H}_B'$. Here $\hat{H}'_0 = \hat{H}_J + \hat{H}_L$ with
\begin{equation}
\hat{H}_J = \frac{\hat{Q}^2}{2C} - E_J \cos \hat{\theta} \mbox{ ; }\hat{H}_L =
\left(\frac{\hbar}{2e}\right)^2\frac{(\phi_b - \hat{\phi}_R - \hat{\theta})^2}{2 L},
\label{single_junction}
\end{equation}
where $\hat{\theta}$ is the phase difference across the junction, conjugate to the charge
$\hat{Q}$, such that $[\hat{Q},\hat{\theta}] = -2ie$. The combination $\phi_b -
\hat{\phi}_R - \hat{\theta}$ is the phase difference across the inductance. It contains
the total phase difference across the circuit $\phi_b$, which is an external parameter
determined by the applied voltage bias $V_b$, such that $\dot{\phi}_b = 2eV_b/\hbar$. The
operator $\hat{\phi}_R$ is the phase difference across the resistor. We express it in
terms of the oscillator bath positions $\hat{\Xi}_i$ according to the relation
\begin{equation}
\hat{\phi}_R = \sum _i \lambda _i \hat{\Xi}_i,
\end{equation}
with coupling constants $\lambda_i$. The dynamics of the bath degrees of freedom,
accounting for dissipation due to the resistor, is governed by the Hamiltonian
$\hat{H}_B'$
\begin{equation}
\hat{H}_B' = \sum_{i=1}^\infty \frac{\hat{\Pi}_i^2}{2} + \frac{1}{2}\tilde{\omega}_i^2
\hat{\Xi}_i^2,
\end{equation}
where $\hat{\Pi}_i$ are the bath momenta conjugate to $\hat{\Xi}_i$; $\tilde{\omega}_i$
are the bath frequencies.

\begin{figure}
\centering
\includegraphics[width=.75\columnwidth,angle=0]{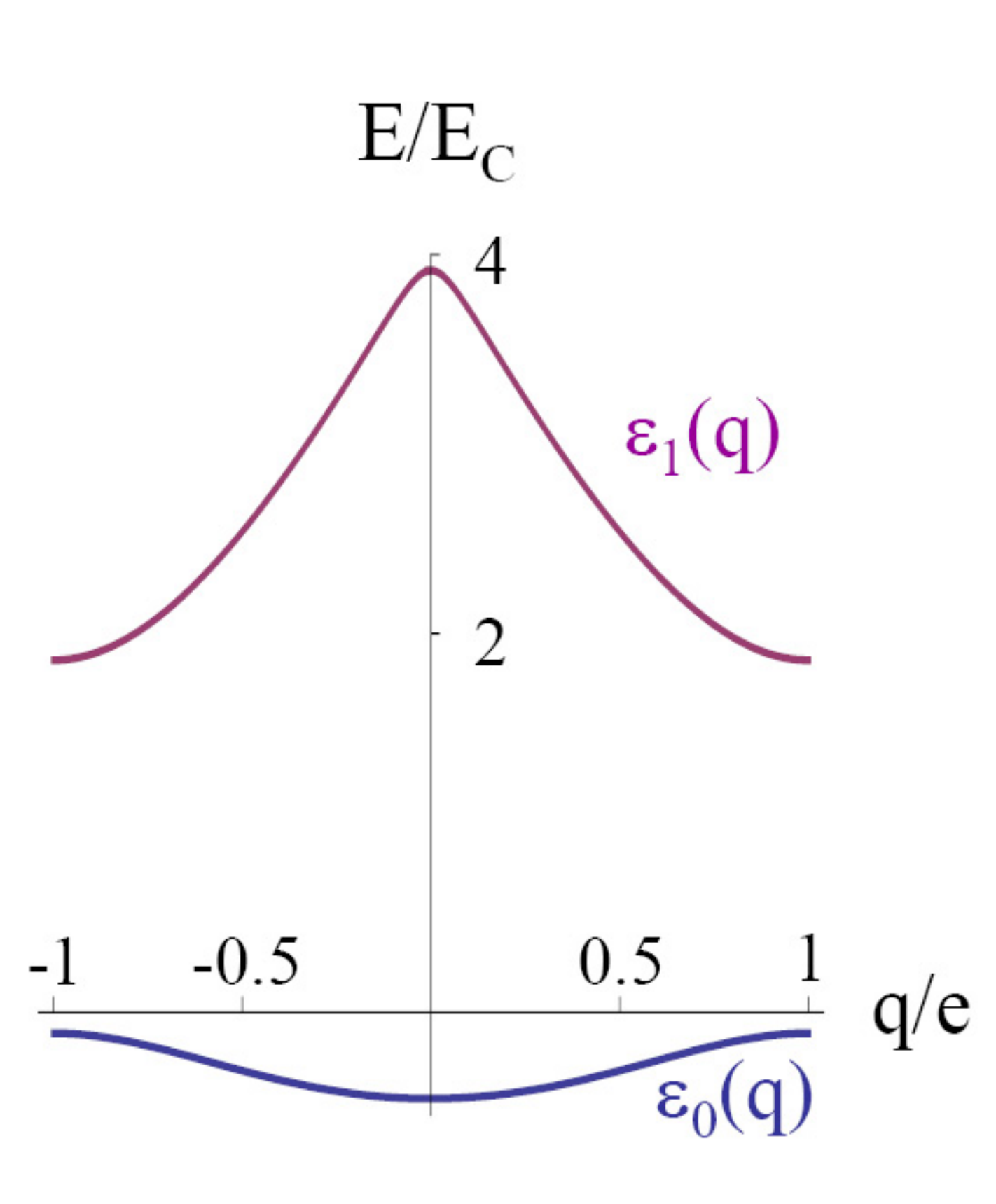}
\caption{Lowest two Bloch bands of the spectrum of Hamiltonian $\hat{H}_J$, taking the
ratio $E_J/E_C = 2$. Note that the bands are separated by an energy of the order of the
plasma frequency, $\hbar \omega_p = 4 E_C$.} \label{Blochband}
\end{figure}

Let us first consider the junction Hamiltonian $\hat{H}_J$, which describes a particle of
mass $C$ moving in a periodic potential. The spectrum of this Hamiltonian consists of
Bloch bands~\cite{Averin_85,Likharev_85}, see Fig.~\ref{Blochband}. We can use the so-called
quasi-charge representation~\cite{Likharev_85,Landau} and write $\hat{H}_J =
\epsilon_0(\hat{q})$, assuming the junction's dynamics to be restricted to the lowest Bloch
band $\epsilon_0(q)$. The commutation relation of the quasi-charge operator $\hat{q}$ and
the phase $\hat{\theta}$ reads $[\hat{q},\hat{\theta}] = -2ie$. The physical meaning of
the quasi-charge becomes clear if we consider its dynamics, which is governed by the
equation of motion
\begin{equation}
\dot{\hat{q}} = \frac{\hbar}{2eL}(\phi_b-\hat{\phi}_R-\hat{\theta})=
\frac{2e}{\hbar}\frac{d \hat{H}'}{d\phi_b}. \label{current}
\end{equation}
Since $\phi_b$ is the externally fixed phase drop over the {\em entire} circuit, the last
term on the right-hand-side corresponds by definition to the {\em total} current flowing
through the circuit. We next calculate $\dot{\hat{\theta}} =
i[\hat{H}',\hat{\theta}]/\hbar$ and find
\begin{equation}
\dot{\hat{\theta}} =  \frac{2e}{\hbar}  \frac{\partial \epsilon_0(q)}{\partial q},
\end{equation}
provided that interband transitions can be ignored~\cite{Landau}. Then the operator
$\partial \epsilon_0(q)/\partial q$ corresponds to the voltage drop over the junction. A
straightforward calculation of $\ddot{\hat{q}}$ now yields the equation of motion for the
quasi-charge. We find
\begin{equation}
L \ddot{\hat{q}} + \partial \epsilon_0(q)/\partial q  = V_b + \delta\hat{V}, \label{JCL}
\end{equation}
where the operator
\begin{equation}
\delta \hat{V} =  -\frac{\hbar}{2e}\sum_{i=1}^\infty \lambda_i \dot{\hat{\Xi}}_i.
\label{dV}
\end{equation}
The bath positions satisfy the equation of motion
\begin{equation}
\ddot{\hat{\Xi}}_i + \tilde{\omega}_i^2 \hat{\Xi}_i = \lambda_i \hbar \dot{\hat{q}}/2e.
\label{X_i}
\end{equation}
Direct integration of Eq.~(\ref{X_i}) yields the sum of the homogeneous and the
particular solution, substitution into Eq.~(\ref{dV}) yields
\begin{equation}
\delta \hat{V} = \hat{v}(t) - \int \limits_0 ^t Z(t-t') \dot{\hat{q}}(t').
\end{equation}
Similar to the situation discussed above for the operator $\delta \hat{I}$, the first
term $\hat{v}(t)$ is the voltage noise related to the uncertainty on the level of the
initial conditions for the homogeneous solution of Eq.~(\ref{X_i}). The second term is
related to the particular solution of Eq.~(\ref{X_i}) and describes the response of the
bath to an applied current $\dot{\hat{q}}$ through the retarded impedance $Z(t)$. The
Fourier transform of the latter is
\begin{equation}
Z(\omega) = \left(\frac{\hbar}{2e}\right)^2\sum _{i=1}^\infty \lambda_i^2\frac{i
\omega}{(\omega +i \eta)^2 - \tilde{\omega}_i^2}.
\end{equation}
If we choose the bath parameters $\lambda_i$ and $\tilde{\omega}_i$ such that
\begin{equation}
\Re \mbox{e}[Z(\omega)] =  \pi \left(\frac{\hbar}{2e}\right)^2 \sum \limits_{i=1}^\infty
\lambda_i^2 \tilde{\omega}_i \delta(\omega^2 - \tilde{\omega}_i^2) = R, \label{bath}
\end{equation}
the bath's response is ohmic corresponding to that of a resistance $R$. Assuming the bath's initial
state to be a canonical equilibrium one, we find that the voltage noise is characterized by the two lowest cumulants,
$\langle \hat{v}\rangle =0$ and $\langle \{\hat{v} (t), \hat{v} (0)\}\rangle/2$, the latter satisfying
the fluctuation-dissipation theorem,
\begin{eqnarray}
S_v(\omega) &\equiv& \int dt e^{i \omega t} \langle \{\hat{v} (t), \hat{v} (0)\}\rangle/2 \nonumber \\
 &=& \hbar \omega \Re \mbox{e}[Z(\omega)] \coth (\hbar \omega/2k_B T) \nonumber \\
 &=& \hbar \omega R \coth (\hbar \omega/2 k_B T).
 \label{symspecfuncv}
\end{eqnarray}

Now consider the limit $E_J \gg E_C$, where $\epsilon_0(q) = - \Delta_0 \cos \pi q/e$
corresponds to a purely sinusoidal band in quasi-charge representation with a bandwidth
given by~\cite{Likharev_85}
\begin{equation}
\Delta_0 = 16 \sqrt{E_J E_C/\pi} (E_J/2E_C)^{1/4} e^{-\sqrt{8 E_J/E_C}}.
\label{D0}
\end{equation}
Then, together with the choice (\ref{bath}) for the oscillator bath
parameters, Eq.~(\ref{JCL}) takes the form
\begin{equation}
L \ddot{\hat{q}} +  R \dot{\hat{q}}+ V_c \sin\pi q/e  = V_b + \hat{v}, \label{CJRL}
\end{equation}
where $V_c = \pi \Delta_0/e$ is the critical voltage. If both sides of Eq.~(\ref{CJRL})
are divided by $V_c$, we obtain the dimensionless form
\begin{equation}
d^2 \hat{\bar{q}}/d\tau'^2 +  \rho d\hat{\bar{q}}/d\tau'+ \sin \hat{\bar{q}}  = \bar{V}_b
+ \hat{\bar{v}} \label{CJRLdimless}.
\end{equation}
Here, $\tau' = \omega_c t$ with $\omega_c = (\pi V_c/eL)^{1/2}$ dual to the plasma
frequency, $\hat{\bar{q}} = \pi \hat{q}/e$, the dimensionless resistance $\rho = R(e/\pi
V_c L)^{1/2}$, $\bar{V}_b = V_b/V_c$ and $\hat{\bar{v}}=\hat{v}/V_c$.

Comparing the quantum Langevin equation Eq.~(\ref{CJRL}) with the corresponding one for
the phase dynamics, Eq.~(\ref{RCSJ}), we conclude that {\em they are exactly dual to each
other}. In other words, Eqs.~(\ref{CJRL}) and (\ref{RCSJ}) map onto each other when
exchanging the role of quasi-charge $\hat{q}$ and phase $\hat{\phi}$, such that $\pi
\hat{q}/e \to \hat{\phi}$, accompanied by the duality transformations $e/\pi \to
\hbar/2e$, $L \to C$, $R \to 1/R$, and $V \to I$. This establishes the exact duality
between the two circuits, Fig.~\ref{JJ}a and b.

Equation (\ref{CJRL}) has a simple physical interpretation in the classical limit: it
describes a fictitious charge-particle with inertia $L$, provided by the inductor. The
sum of the applied bias voltage $V_b$ and the resistor-induced noise $v$ drops over the
series configuration formed by the junction, the inductor and the resistor. The
charge-particle moves in a tilted washboard potential $U(q) = -\delta_0 \cos \pi q/e -
V_b q$ and experiences velocity-proportional damping $R \dot{q}$.

As for its dual counterpart, the classical interpretation of the operator equation
(\ref{CJRL}) hinges on the replacement of the operators $\hat{q}$ and $\hat{v}$ by their
respective expectation values $q$ and $v$. This means in particular that we must be
allowed to replace $\langle \sin \pi \hat{q}/e \rangle$ by $\sin \langle \pi \hat{q}/e
\rangle = \sin \pi q/e$. For this to be correct, the uncertainty in the charge $\delta q$
must be small compared to the elementary charge $e$. The classical version of
(\ref{CJRL}) then describes the motion of a narrow wave packet of width $\delta q \ll e$.
In phase representation this is consistent with the realization of an extended Bloch
state.

As for the case of phase dynamics, it is useful to distinguish two situations, according
the value of the parameter $\rho$ in Eq.~(\ref{CJRLdimless}): overdamped charge dynamics,
corresponding to $\rho
> 1$ and underdamped charge dynamics, $\rho <1$.

In the overdamped case, the charge dynamics is always classical as damping times are
naturally short, of the order of $L/R$. Note that $L$ here constitutes an additional
element in addition to the junction, unlike the case of phase dynamics discussed above
where the capacitance $C$ is a property intrinsic to the junction. Overdamped charge
dynamics is generally achieved in a high-resistance environment, which favors charge
localization. Classical charge dynamics also requires that the environmental noise
operator $\hat{v}$ can be treated classically~\cite{Corlevi_06}. This implies working in
the high-temperature limit  $k_B T > \hbar R/L$, such that $\hat{v}$ can be replaced by a
c-number with $\langle v \rangle =0$ and $\langle v(t) v(0)\rangle = 2 k_BT R \delta(t)$
in accordance with Eq.~(\ref{symspecfuncv}). The resulting classical Langevin equation
can be analyzed through the corresponding Fokker-Planck
equation~\cite{Kampen,Beloborodov_03}.

In the underdamped case, damping times are long and wave packet spreading becomes
important. This issue is particularly important when the charge-particle is in the
running state corresponding to a finite current through the junction. In the absence of
damping, the uncertainty $\delta q$ can be kept within limits by applying classical
time-dependent external forces. The spreading of wave packets is governed by a rate
proportional to the kinetic energy. For the fictitious charge particle this corresponds
to the inductive energy $E_L = \Phi_0^2/2 L$, which is in fact the energy scale relevant
for phase localization related to the Josephson effect. We require the time-dependence of
the external force to be fast on the scale $\hbar/E_L$. Physically this means that we
work under conditions where phase slip events are not suppressed. For example, in the
case of microwaves discussed in Section~\ref{MW_effects} below, this implies their frequency to be larger than $E_L/\hbar$.

We conclude this section by summarizing the conditions under which the above exact
duality is obtained. First of all, we assume the quasi-charge dynamics of the
voltage-biased circuit Fig.~\ref{JJ}b to be determined by the lowest Bloch band only. The
lowest band is separated from the next one by the plasma frequency $\hbar \omega_p =
\sqrt{8 E_J E_C}$. Since we do not consider here inter-band transitions, we thus assume
all energies to be smaller than $\omega_p$. Second, in order for the Bloch band to be
purely sinusoidal, we need to impose the condition $E_J \gg E_C$ for the voltage-biased
junction of Fig.~\ref{JJ}b.

\section{Josephson junction chain}
\label{arrays}

In the previous section we have found that a voltage-biased Josephson junction with a
large ratio $E_J/E_C$ in series with an inductance constitutes a phase-slip element. We
also saw that the observation of well-defined quasi-charge dynamics requires
charge-fluctuations very much smaller than a single charge $e$, a condition that can be
obtained by a large inductance near the Josephson junction. Experimentally it is not
so simple to realize large \emph{magnetic} inductances very close to the sample. An
alternative is to fabricate a large \emph{kinetic} inductance using a superconductor, and
in particular a Josephson junction chain\cite{Watanabe_01,Corlevi_06,Devoret_09}. It seems therefore quite
natural to analyze the possibility to realize a phase-slip junction from a Josephson
junction chain, which we investigate in this section. The central idea is that the phase-slip itself occurs on only one of
the junctions of the chain; the phases on the other junctions just perform small
Josephson oscillations, thereby providing the necessary inductance. In order to
demonstrate this idea, we will closely follow the paper by Matveev et
al.~\cite{Matveev_PRL02}, who studied the low-energy properties of Josephson junction
chains in terms of quantum phase slips.

\begin{figure}
\includegraphics[width=\columnwidth,angle=0]{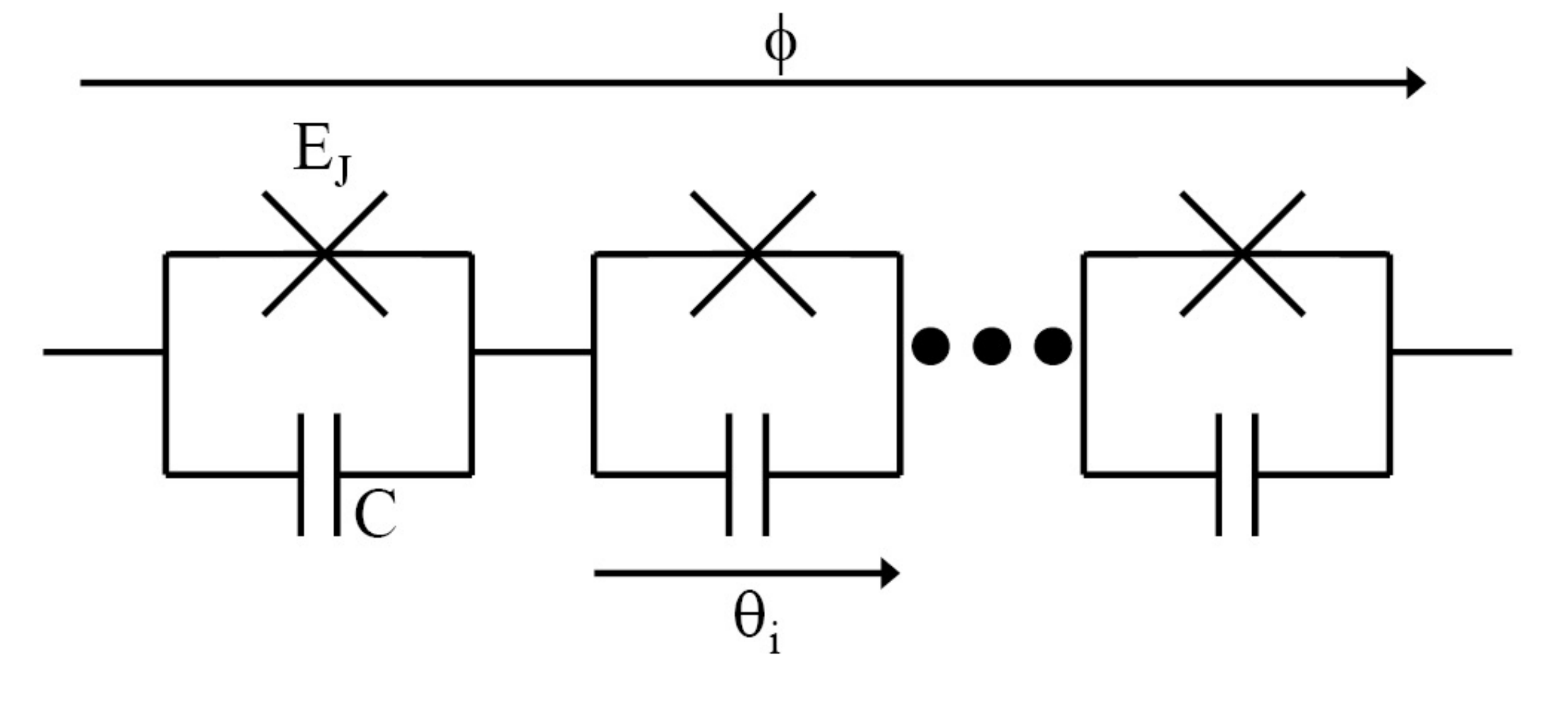}
\caption{Josephson junction chain.}\label{array}
\end{figure}

We start our analysis by considering the Josephson junction chain depicted in
Fig.~\ref{array}. It is a series arrangement of $N$ identical junctions, each with
Josephson energy $E_J = \hbar I_c/2e$ and charging energy $E_C = e^2/2C$. Let $Q_i$ be
the charge on the $i$-th junction and $\theta_i$ the conjugate phase difference. We
consider the nearest-neighbor-capacitance limit and assume the chain to be subjected to
an external phase $\phi$. The Hamiltonian can then be written as
\begin{equation}
H_\mathrm{ar} = \sum _{i=1}^N [4 E_C (Q_i/2e)^2 + E_J(1 - \cos \theta_i)] \mbox{ ;
}\sum_{i=1}^N \theta_i = \phi.
\end{equation}

Ignoring the charging energy for the moment, we find the classical ground state
configuration of the chain's phases $\theta_i$ by minimizing the Josephson coupling
energy, thereby satisfying the constraint. The corresponding configuration can be easily
found in the limit $N \gg 1$ and is given by $\theta_i = \phi/N$. The resulting Josephson
energy hence reads $E_0 = E_J \phi^2 /2N$. This is the inductive energy stored in the
chain; the corresponding effective inductance is given by $L_\mathrm{ar} = \hbar N/2e
I_c$.

Now consider a phase slip event occurring on one of the junctions, say the $j$th
junction, such that $\theta_j \to \theta_j + 2\pi$. Due to its periodicity as a function
of $\theta_j$, the Josephson energy of the junction $j$ does not change. However, the
constraint $\sum _i \theta_i = \phi$ is violated after such a phase-slip event. In order
to accommodate the phase-slip without violating the constraint, the phase differences
$\theta_i$ over the other junctions change slightly, from $\phi/N$ to $(\phi-2 \pi)/N$.
Correspondingly, the Josephson energy of the chain changes from $E_0 = E_J \phi^2 /2N$ to
$E_1 = E_J (\phi - 2\pi )^2 /2N$. In a similar way one shows that the classical energy
needed to accommodate $m$ phase-slips without violating the constraint is given by $E_m =
E_J (\phi - 2\pi m)^2 /2N$. We thus conclude that the ground state of the chain generally
is one that contains a fixed number of phase slips for almost any value of the external
phase $\phi$, except for the special values $\phi = \pi (2 m +1)$ where the energies
$E_m$ and $E_{m+1}$ are degenerate. Quantum fluctuations induced by the small but finite
charging energy $E_C$ lift this degeneracy: they give rise to a non-vanishing amplitude
$\Delta_0$ for a phase-slip event to occur. We denote the state of the chain with $m$
phase slips by $|m\rangle$. Taking into account the fact that a phase slip can take place
on any of the $N$ junctions, we can write the total Hamiltonian for the chain as
\begin{equation}
\hat{H}_\mathrm{ar} = \frac{E_J}{2N}(2 \pi \hat{m} - \phi)^2
- \frac{N \Delta_0}{2} \sum _m
[|m+1\rangle\langle m| + h.c.].
\end{equation}

Next introduce the operator $\hat{q}$, conjugate to the phase-slip number $\hat{m}$.
Specifically, $[\hat{q},\hat{m}] = -ie/\pi$ such that the operator $e^{i \pi \hat{q}/e}$
is a raising operator with $e^{i \pi \hat{q}/e} |m\rangle = |m+1\rangle$. Using this
representation, the Hamiltonian becomes
\begin{equation}
\hat{H}_\mathrm{ar} = (E_J/2N) (2 \pi \hat{m} - \phi)^2 - N \Delta_0 \cos \pi \hat{q}/e. \label{hamarr}
\end{equation}
A physical interpretation of the operator $\hat{q}$ can be obtained by calculating
$\dot{\hat{q}}$; the result reads $\dot{\hat{q}} = (2e/\hbar) d\hat{H}/d\phi$, which by
definition is the operator corresponding to the total current through the chain. Hence,
in analogy with the result (\ref{current}), $\hat{q}$ can be interpreted as the global
charge of the chain. It is easy to see that Hamiltonian Eq.~(\ref{hamarr}) has the same
form as the quasi-charge representation of Hamiltonian $\hat{H}_0'$ for a single junction
in series with an inductor, presented in the previous section. It is therefore
straightforward to analyze the case of a voltage-biased chain embedded in a resistive
series environment, repeating the steps presented in Section~\ref{juncind}. One finds
that the quasi-charge dynamics is governed by the equation
\begin{equation}
L_\mathrm{ar} \ddot{\hat{q}} +  R \dot{\hat{q}}+ V_{c,\mathrm{ar}} \sin\pi \hat{q}/e  =
V_b + \hat{v}, \label{CJRLarray}
\end{equation}
where $V_{c,\mathrm{ar}} = N V_c$ and we used the fact that $\dot{\phi}_b = 2eV_b/\hbar$.
This result can also be presented in the dimensionless form~(\ref{CJRLdimless}), with
$\omega_{c,\mathrm{ar}} = (\pi V_{c,\mathrm{ar}}/eL_\mathrm{ar})^{1/2} = (2 \pi
V_{c}I_c/\hbar)^{1/2}$ and $\rho = R(e/\pi V_{c,\mathrm{ar}} L_\mathrm{ar})^{1/2} =
(R/N)(2e^2 I_c/\pi \hbar V_c)^{1/2}$. This result reflects the intuitive argument
discussed at the beginning of this section: the inertia $L_\mathrm{ar}$ of the charge
dynamics is provided by the chain itself. Moreover, it is given by $N$ times the
nonlinear inductance $\hbar/2eI_c$ of a single junction in the chain. This means that
$L_\mathrm{ar}$ can be tuned in principle, either by tuning $N$ or by using SQUID loops
instead of single junctions as in~\cite{Corlevi_06}, such that $I_c$ can be tuned with a
magnetic flux. We also note that the critical voltage of a Josephson junction chain is
$N$ times larger than the one of a single Josephson junction. This is relevant for the
discussion in the next section, where we will analyze the $I$-$V$ characteristic of a
phase-slip junction under microwave irradiation. As the width of the appearing current
steps  scales with the critical voltage of the phase-slip junction, a Josephson junction
chain has necessarily larger current-steps that are as a consequence more robust against
voltage noise.

\section{Phase-slip junction under the influence of microwave irradiation}
\label{MW_effects}

\begin{figure}
\centering
\includegraphics[width=\columnwidth,angle=0]{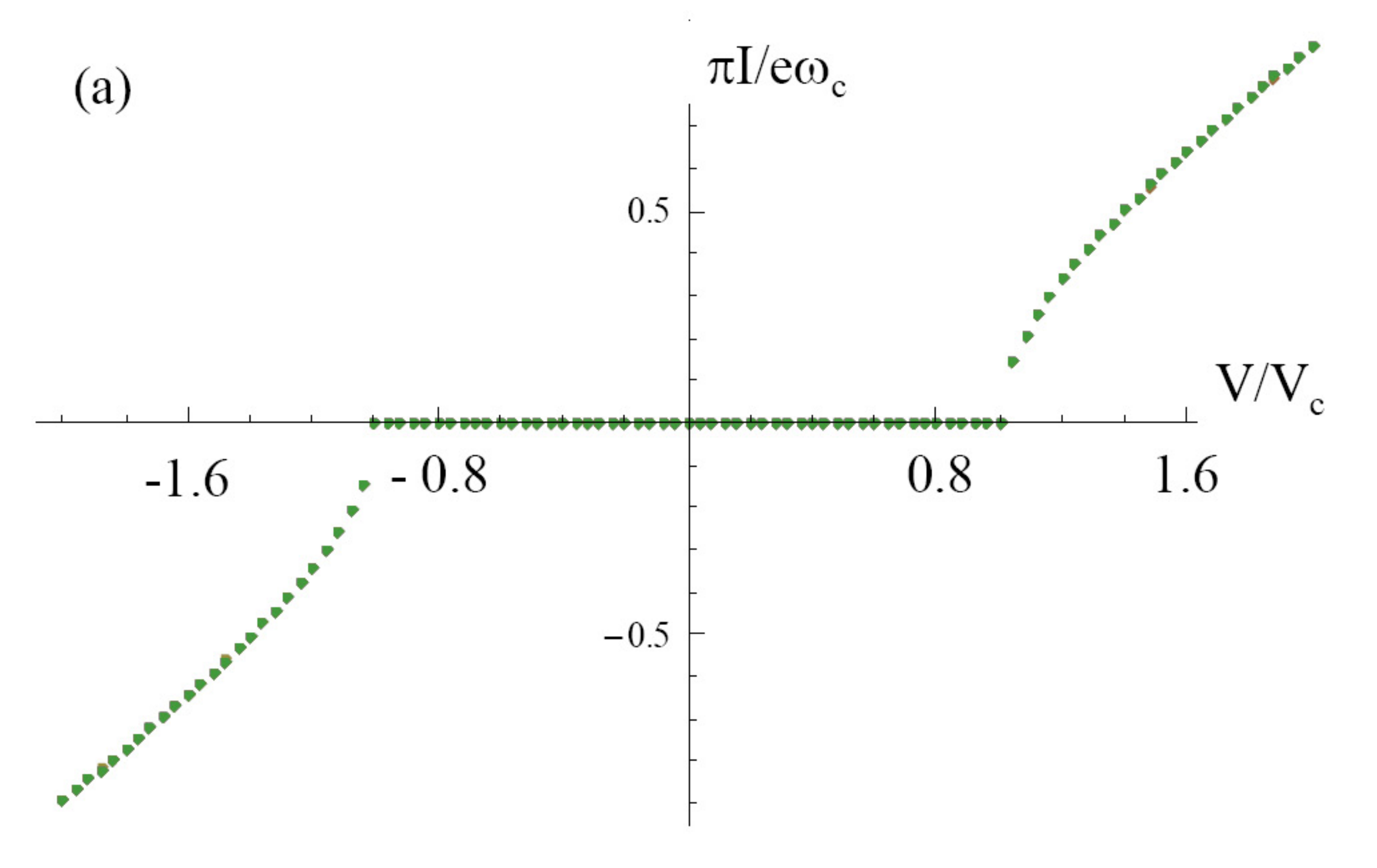}
\includegraphics[width=\columnwidth,angle=0]{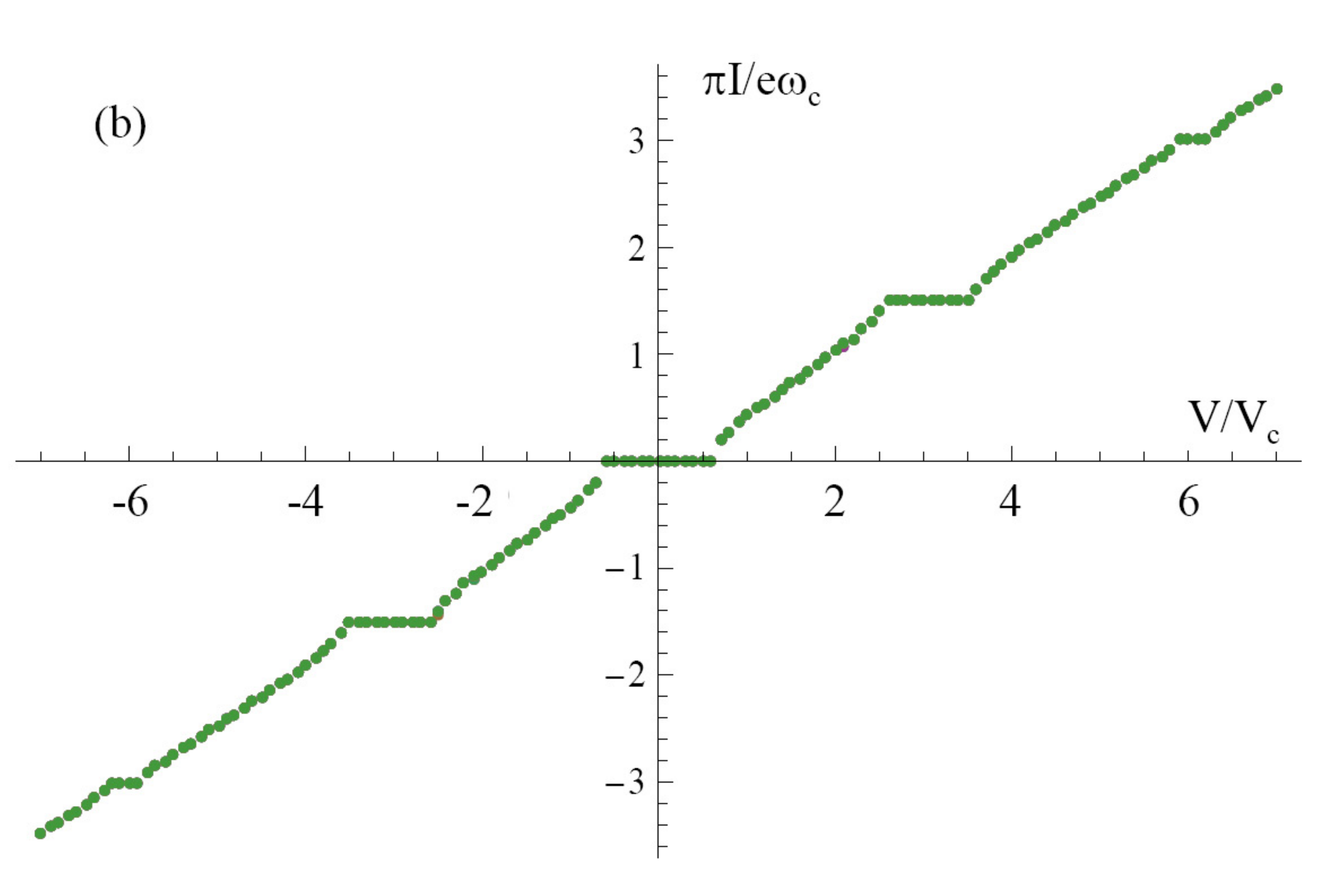}
\caption{Current-voltage characteristics for an overdamped phase-slip junction with
$\rho =2$, (a) without MW irradiation, (b) under MW-irradiation with amplitude $V_\mathrm{MW} =
5 V_c$ and frequency $\omega_\mathrm{MW} = 1.5 \omega_c$.} \label{overdamped}
\end{figure}

\begin{figure}
\centering
\includegraphics[width=\columnwidth,angle=0]{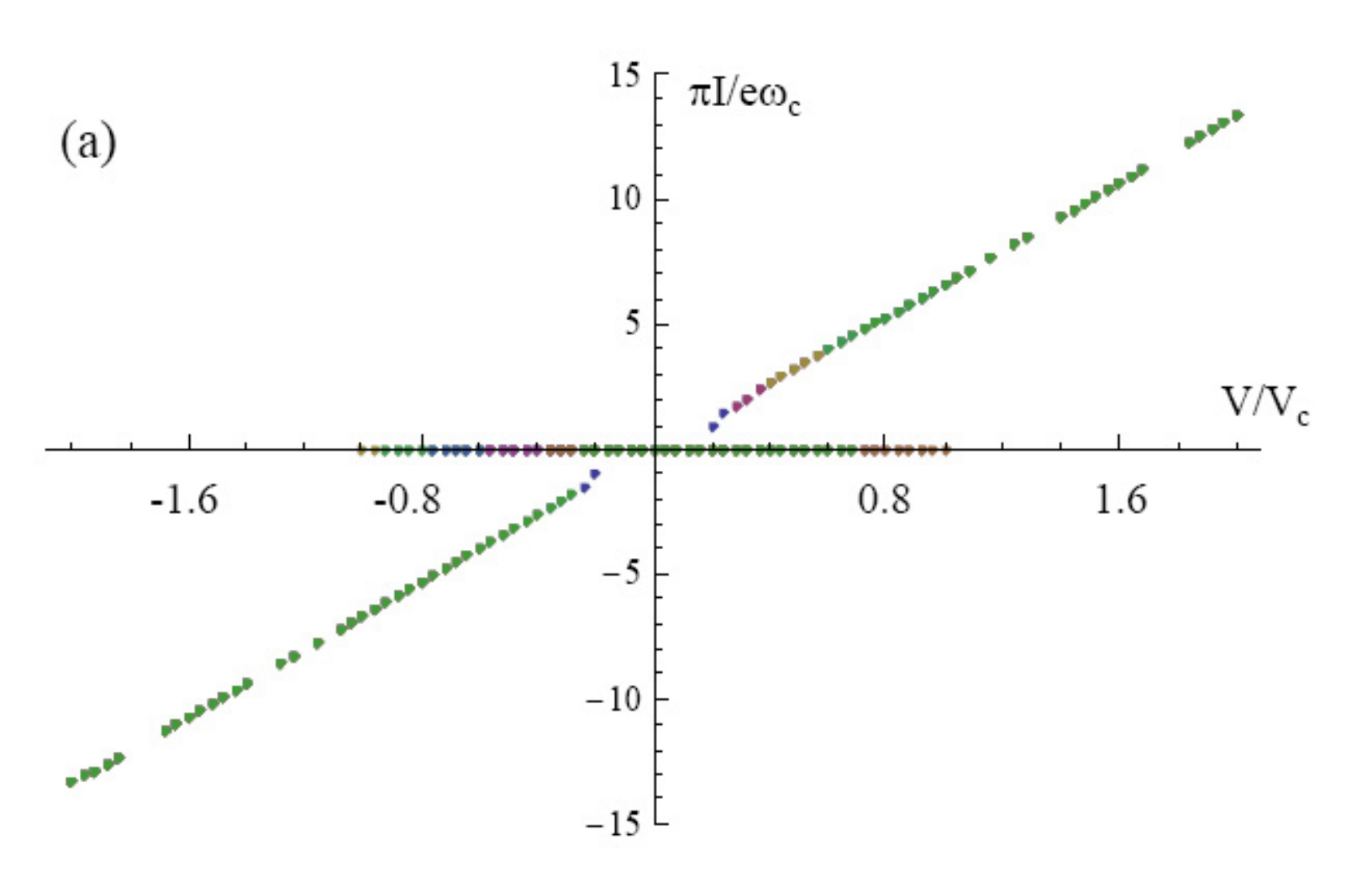}
\includegraphics[width=\columnwidth,angle=0]{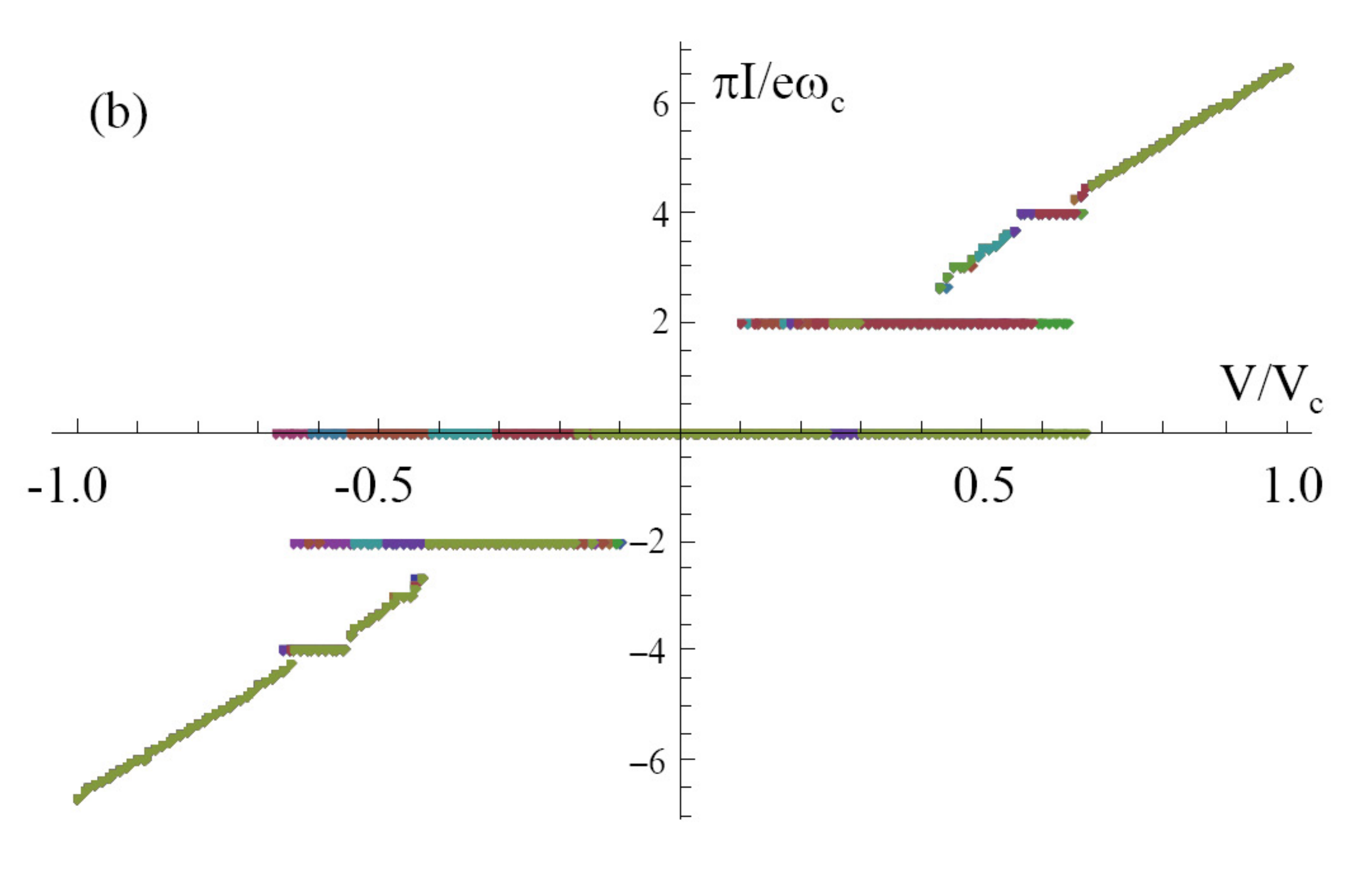}
\caption{Current-voltage characteristics for an underdamperd phase-slip junction with
$\rho =0.15$, (a) without MW irradiation, (b) under MW-irradiation with amplitude $V_\mathrm{MW} = 3
V_c$ and frequency $\omega_\mathrm{MW} = 2 \omega_c$.} \label{underdamped}
\end{figure}

As it was already mentioned in the Introduction, it is of interest to study the
behavior of phase-slip junctions under the influence of microwaves. In view of
duality, we expect the current-voltage characteristics to exhibit steps at
well-defined values of the current that are multiples of the microwave frequency,
so-called current Shapiro steps. In order to demonstrate this, we have numerically
integrated Eq.~(\ref{CJRL}) in the classical limit. We set $V_b (t) = V +V_\mathrm{MW} \sin
(\omega_\mathrm{MW}t)$, and ignore the effect of fluctuations. The results are plotted
in Figs.~\ref{overdamped} and \ref{underdamped} for two choices of the dimensionless
damping parameter $\rho$, corresponding to the overdamped and underdamped limit,
respectively.

Figures~\ref{overdamped}a and \ref{underdamped}a show the current-voltage
characteristics in the absence of microwaves, $V_\mathrm{MW}=0$. In the overdamped case,
Fig.~\ref{overdamped}a, the current remains zero as long as the voltage $V$ is
smaller than the critical voltage $V_c$: the phase-slip junction is in the Coulomb
blockade regime. Beyond $V_c$, the current rapidly increases until it reaches the
linear asymptote $I \sim V/R$, where the junction is in the superconducting state
and the voltage drops entirely over the resistor. This characteristic corresponds in fact to
the so-called Bloch nose, which has been
studied in the experiment~\cite{Corlevi_06} in the presence of thermal fluctuations. It was found that such fluctuations
induce a smooth interpolation between the Coulomb blockade and
the superconducting branch, in agreement with theory~\cite{Beloborodov_03}. In the underdamped situation, Fig.~\ref{underdamped}a,
the two branches in the characteristic co-exist in the Coulomb blockade region. Since both branches are accessible, we
generally expect to find hysteretic behavior of the phase-slip junction. To the best
of our knowledge, this limit has not yet been accessed experimentally. The behavior
shown in Figs.~\ref{overdamped}a and \ref{underdamped}a have a dual counterpart in the
usual RCSJ-model, where both the overdamped and the underdamped limit have been
studied in experiments~\cite{Steinbach_01,Tinkham}.

If microwaves are applied, steps appear at well-defined values of the current that are
multiples of the applied MW frequency, as can be seen in Figs.~\ref{overdamped}b and
\ref{underdamped}b. The steps are clearly visible; they are integrated within the
overall $I$-$V$ characteristic, the average slope of which remains determined by the
resistance $R$. In the overdamped case, there is a one-to-one correspondence between
current and voltage, as for the case without MW. In fact, the shape of each step in
Fig.~\ref{overdamped}b appears to be a replica of the characteristic in the absence of
MW, Fig.~\ref{overdamped}a. The one-to-one correspondence between voltage and current
found in the overdamped limit is lost in the underdamped case: various current steps
appear within the same voltage interval. In order to obtain the result shown in Fig.~\ref{underdamped}b,
Eq.~(\ref{CJRL}) had to be integrated for a range of initial conditions on the quasi-charge for each value
of the DC voltage $V$. Note that the resistive branch is absent in the regions of overlapping
steps; this is of interest as it possibly makes it easier to lock on a given step in
the experiment.

\section{Conclusions and outlook}
\label{outlook}

We have studied two circuits, a single Josephson junction in an inductive-resistive environment and a Josephson junction chain, in view of the realization of the dual of the Josephson effect. In both cases we elucidate the importance of the role of the inductance in order to reduce charge fluctuations. Duality between these two circuits and a single Josephson junction is valid in the quantum and classical regime. Here, we derived $I(V)$ characteristics in the classical quasi-charge regime. In case of larger quasicharge fluctuations going beyong the classical regime, perturbation theory can be applied in analogy to the $P(E)$ theory in Josephson junctions governed by phase-dynamics~\cite{IngoldNazarov,Averin_90,Zazunov_08}.

Until now there are only a few experiments dealing with quasi-charge dynamics~\cite{Kuzmin_91, Watanabe_01, Nguyen_07, Corlevi_06}. We believe
that both circuits are experimentally feasible and they are of particular interest in
terms of the realization of current Shapiro steps. The successful realization of such an
experiment would link the frequency to the current by a quantum electrical recipe and
close the metrological triangle. Ultimately, the quantum metrological triangle experiment
would enable a consistency check of the fundamental constants of nature, the electron
charge $e$ and Planck's constant $h$, and as a consequence link the kilogram (the only unit
still defined on the basis of one artefact, prototype of the mass kept in metrology
institutions), to the Planck constant. The circuit based on a single Josephson chain
seems from our point of view the most promising as it enables to realize large current
steps.

\acknowledgments We wish to thank O. Buisson, N. Didier, L. Glazman, D. Haviland, L. Kuzmin, and A. Zaikin for valuable discussions. Financial support from the European Community (STREP MIDAS) and the French National Research Agency (ANR QUANTJO) is gratefully acknowledged. FH thanks the ICTP Trieste (Italy), where part of the work described here has been performed, for hospitality.

\end{document}